\documentclass[sigconf]{acmart}
\usepackage{enumitem}
\AtBeginDocument{%
  }
\copyrightyear{2026}
\acmYear{2026}
\setcopyright{cc}
\setcctype{by}
\acmConference[GLSVLSI '26]{Great Lakes Symposium on VLSI 2026}{June 22--24, 2026}{Canandaigua, NY, USA}
\acmBooktitle{Great Lakes Symposium on VLSI 2026 (GLSVLSI '26), June 22--24, 2026, Canandaigua, NY, USA}
\acmDOI{10.1145/3787109.3815245}
\acmISBN{979-8-4007-2431-2/2026/06}
\begin{document}
\title{Synthesis-in-the-Loop Evaluation of LLMs for RTL Generation: Quality, Reliability, and Failure Modes}
\titlenote{Accepted at the Great Lakes Symposium on VLSI 2026 (GLSVLSI '26), June 22--24, 2026, Canandaigua, NY, USA. This is the authors' preprint; the published version is available at \url{https://doi.org/10.1145/3787109.3815245}. Code: \url{https://github.com/owenfucell/SynthAgentic}. Evaluation data: \url{https://huggingface.co/datasets/KSU-HW-SEC/SynthAgentic-Eval}.}

\author{Weimin Fu}
\email{weiminf@ksu.edu}
\affiliation{%
  \institution{Kansas State University}
  \city{Manhattan}
  \state{Kansas}
  \country{USA}
}

\author{Zeng Wang}
\email{zw3464@nyu.edu}
\affiliation{%
  \institution{NYU Tandon School of Engineering}
  \city{Brooklyn}
  \state{New York}
  \country{USA}
}

\author{Minghao Shao}
\email{shao.minghao@nyu.edu}
\affiliation{%
  \institution{NYU Tandon School of Engineering}
  \city{Brooklyn}
  \state{New York}
  \country{USA}
}

\author{Ramesh Karri}
\email{rkarri@nyu.edu}
\affiliation{%
  \institution{NYU Tandon School of Engineering}
  \city{Brooklyn}
  \state{New York}
  \country{USA}
}

\author{Muhammad Shafique}
\email{muhammad.shafique@nyu.edu}
\affiliation{%
  \institution{NYU Abu Dhabi}
  \city{Abu Dhabi}
  \country{UAE}
}

\author{Johann Knechtel}
\email{johann@nyu.edu}
\affiliation{%
  \institution{NYU Abu Dhabi}
  \city{Abu Dhabi}
  \country{UAE}
}

\author{Ozgur Sinanoglu}
\email{ozgursin@nyu.edu}
\affiliation{%
  \institution{NYU Abu Dhabi}
  \city{Abu Dhabi}
  \country{UAE}
}

\author{Xiaolong Guo}
\email{guoxiaolong@ksu.edu}
\affiliation{%
  \institution{Kansas State University}
  \city{Manhattan}
  \state{Kansas}
  \country{USA}
}

\renewcommand{\shortauthors}{Fu, Wang, Shao, et al.}
\begin{abstract}
RTL generation is more than code synthesis. Designs must be syntactically valid, synthesizable,  correct,  hardware-efficient. SOTA evaluations stop at functional correctness and do not measure synthesis and implementation quality. This paper evaluates 32 language models on 202 Verilog tasks from VerilogEval and RTLLM using the Hardware Quality Index (HQI) that combines post-synthesis area, delay, and warnings related to expert references in a Nangate45 45\,nm flow.
Three performance regimes emerge: 14 frontier models achieve HQI $>$ 66, led by Gemini-3-Pro at 87.5\% coverage and 85.1 HQI; 15 models cluster 43--66 HQI; 3 are below 43. Gap between best-of-five capability and single-attempt quality spans 3.7--22.1 HQI points, limiting integration into agentic pipelines. A taxonomy of 195 synthesis failures reveals systematic divergence: proprietary models fail late through elaboration errors and synthesis timeout; open models fail early often due to missing module wrappers and non-synthesizable constructs, a pattern consistent with training corpora skewed toward simulation over synthesis-grade RTL. 
\end{abstract}



\begin{teaserfigure}
    \includegraphics[width=1\linewidth]{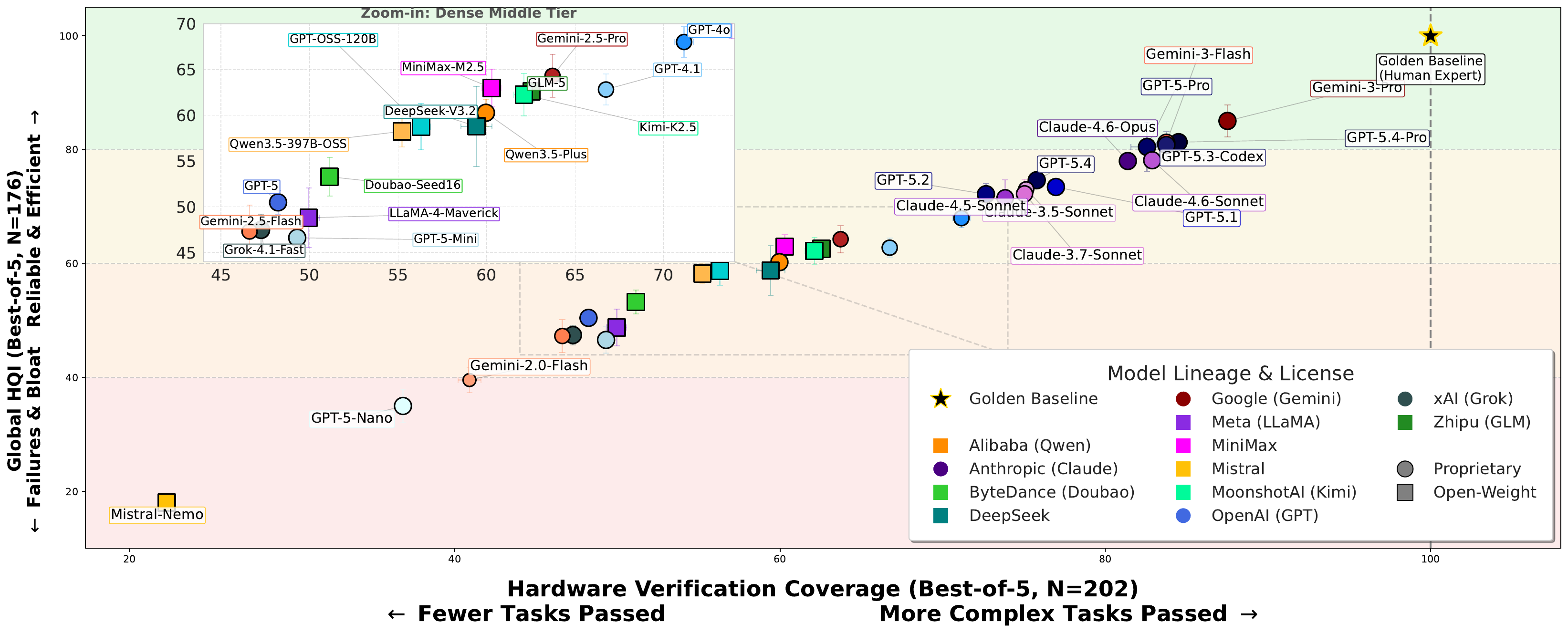}
    \Description{Scatter plot of 32  models on Coverage--Global HQI plane, revealing three tiers. Horizontal axis shows complexity-weighted verification coverage (0--100\%) and the vertical axis shows Global HQI (0--100). Tier-1 models (Global HQI above 66) cluster in the upper right, led by Gemini-3-Pro at 87.5\% coverage and 85.1 HQI. An inset zooms into the dense Tier-2 cluster. Marker shape distinguishes proprietary (circles) from open-weight (squares) models; color encodes provider family.}
    \caption{Coverage--Global HQI  landscape for 32  models under synthesis-in-the-loop evaluation. Three tiers emerge at boundaries identified by k-means clustering: Tier~1 (Global HQI $\geq 66$, 14 models), Tier~2 (43--66, 15 models), and Tier~3 ($<43$, 3 models). Circles are proprietary models; squares are open-weight. Inset magnifies the dense Tier~2 cluster. Exact values are in Table~\ref{tab:main_results}.}
    \label{fig:teaser}
\end{teaserfigure}

\maketitle

\section{Introduction}\label{sec:intro}
Large language models have advanced code generation~\cite{chen2021codex,austin2021mbpp,li2022alphacode}, motivating their application to hardware description languages~\cite{liu2023verilogeval,liu2024openllmrtl,liu2025rtlcoder,wang2024llms}. RTL generation is a more demanding setting than software synthesis: a design must be syntactically valid, synthesizable into gates, functionally correct under simulation, \emph{and} be area- and timing- efficient~\cite{lu2024rtllm,pinckney2025verilogeval}. Software-style evaluation metrics cannot capture them. Register-transfer-level (RTL) design is the primary productivity bottleneck in the chip development cycle: engineering teams must satisfy function correctness, timing closure, and area budgets across every module of a potentially million-gate design, a process consuming months to years per tape-out. Automated RTL generation could compress this cycle by shifting effort from authoring to reviewing, but only if generated designs satisfy the synthesis stack. Passing a simulation testbench is necessary but not sufficient; synthesis-aware quality evaluation assesses whether LLM-generated RTL can substitute for expert written ones. 

Agentic AI amplifies this evaluation challenge. AI-driven automation spans chip placement~\cite{mirhoseini2021graph}, logic synthesis~\cite{hosny2020drills}, RTL generation~\cite{fu2023gpt4aigchip}, and conversational design reaching tapeout~\cite{blocklove2023chipchat}, with autonomous agents orchestrating end-to-end EDA workflows~\cite{wu2024chateda}. In these pipelines, the model's \emph{first} attempt propagates downstream: $k$ candidates at each of $n$ stages would yield a $k^{n}$-leaf tree, making multi-sample strategies combinatorially intractable. First-attempt expected quality hence is a key metric, shifting the focus from  pass@$k$ feasibility evaluations to deployment-grade assessment of what quality one should \emph{expect} from a single call. Quantifying the gap between best-of-$k$ promise and pass@$1$ reality under synthesis-aware criteria is key to integrating LLMs into agentic design flows.

Existing HDL benchmarks evaluate  syntax validity or functional pass rates~\cite{liu2023verilogeval,thakur2024verigen,liu2024openllmrtl,pinckney2025verilogeval,wang2025vericontaminated}, which do not distinguish designs that merely pass a testbench from those that compile into efficient hardware. Benchmark conclusions can be further distorted by evaluator artifacts~\cite{liu2023humanevalplus,siddiq2024faultinstars,aleithan2025swebench}. The gap is not merely quantitative: a design may pass all assertions yet contain non-synthesizable constructs, or incur 2--5$\times$ area penalties relative to an expert reference. Neither failure mode is detectable by simulation alone, making synthesis-in-the-loop evaluation necessary. This paper evaluates 32 language models on 202 Verilog tasks from VerilogEval and RTLLM, with five independent attempts each, under a staged pipeline: syntax validity via Icarus Verilog, synthesizability via Yosys under the Nangate45 45\,nm library, and functional correctness via testbench simulation. HQI, a 0--100 score, is awarded only to designs clearing all three gates and is then anchored to how closely post-synthesis area, delay, and warning count match an expert golden reference under the same flow. Complexity-weighted Coverage and Global HQI ensure harder multi-module designs contribute proportionally more.

Three key findings emerge. The 32 models group into three capability tiers, with a $4.7\times$ spread in hardware implementation quality between the strongest and weakest. 
Simulation-based pass rates overstate hardware readiness by up to 15 HQI points relative to synthesis-aware scores. Best-of-five capability consistently exceeds single-attempt  quality, exposing a deployment-readiness gap. This limits integration into agentic pipelines where only first attempt propagates. A failure taxonomy reveals that proprietary and open-weight models fail via different mechanisms, with proprietary models exhibiting late elaboration errors and open-weight models exhibiting early structural violations. 
We make three contributions:
\begin{enumerate}[leftmargin=*,noitemsep,topsep=3pt]
\item A synthesis-in-the-loop evaluation pipeline and HQI, a 0--100 metric using post-synthesis area, delay, and warning count to expert golden references under Nangate45 flow, to assess RTL implementation quality beyond pass/fail correctness.
\item An ecosystem-wide empirical characterization of 32 language models on 202 RTL tasks with five independent attempts each, revealing a structured three-tier capability landscape and quantifying the deployment-reliability gap across all tiers.
\item A nine-class synthesis failure taxonomy exposing systematic differences in \emph{how} proprietary and open-weight models fail, with late elaboration errors dominating the former and early structure violations dominating the latter. 
\end{enumerate}

\section{Background}
Large language models (LLMs), transformer-based neural networks pretrained on large text and code corpora and capable of generating code from natural-language prompts, have demonstrated strong performance on software synthesis benchmarks such as HumanEval~\cite{chen2021codex} and MBPP~\cite{austin2021mbpp}. These results have motivated the use of LLMs to hardware description languages, where early work showed that general-purpose models can generate syntactically valid Verilog with reasonable accuracy~\cite{thakur2024verigen, liu2025rtlcoder}. Dedicated HDL benchmarks grade correctness via simulation. VerilogEval~\cite{liu2023verilogeval} grades 155 single-module tasks with Verilator, establishing pass@$k$ (probability that at least one of $k$ independent generation attempts produces a passing solution) as the  primary metric. RTLLM~\cite{lu2024rtllm} and its expansion OpenLLM-RTL~\cite{liu2024openllmrtl} extend coverage to 47 real-world RTL problems, but retain simulation pass rate as the quality criterion. Specialized models have emerged to optimize these metrics: VeriGen~\cite{thakur2024verigen} fine-tunes on open-source Verilog corpora, and RTLCoder~\cite{liu2025rtlcoder} shows that a lightweight model can exceed GPT-3.5 on VerilogEval pass rates. A systematic re-evaluation~\cite{pinckney2025verilogeval} confirms year-over-year progress on simulation metrics but argues that quality-of-results (QoR) measures such as area and timing are necessary for a full picture. The blind spot is consistent: all successors and extensions of these benchmarks use simulation-only evaluation and do not assess whether the designs are synthesizable or hardware-efficient.

We use VerilogEval and RTLLM source HDL benchmarks with published model comparability. They provide a baseline for cross-study comparison. This work is orthogonal to task-set design: rather than expanding the task corpus, it upgrades evaluation criterion from simulation pass/fail to  implementation quality via synthesis-in-the-loop scoring. Logic synthesis transforms RTL into a gate-level netlist and is a key step between design intent and physical implementation. Post-synthesis metrics (area, critical-path delay, and synthesis warnings) are QoR indicators in digital design flows. Open-source tools such as Yosys~\cite{wolf2013yosys} paired with standard-cell libraries (e.g., Nangate45~\cite{nangate45}) yield reproducible, technology-realistic synthesis evaluation. Prior LLM RTL benchmarks do not integrate synthesis-stage QoR into its scoring. HQI metric generalizes to any  RTL benchmark for synthesis-aware quality assessment. 

\section{Methodology}\label{sec:method}
\subsection{Benchmark and Model Set}

The task set has 202 Verilog design problems  from two benchmarks. \textbf{VerilogEval}~\cite{liu2023verilogeval} contributes 155 single-module tasks; each task specifies a fixed top-level interface that the design must wrap correctly. \textbf{RTLLM}~\cite{lu2024rtllm,liu2024openllmrtl} contributes 47 real-world problems including barrel shifters, clock dividers, FSMs, and multi-stage pipelines, often requiring multiple interdependent submodules. All 202 tasks have complexity weights $C_t$ derived from AST dependency-edge counts of their golden reference designs, normalized to $[1, 24]$:
\[C_t = 1 + 23 \cdot \frac{\mathtt{dep\_edges\_ast}(D_t) - e_{\min}}{e_{\max} - e_{\min}}.\]
Of the 202 tasks, 176 have valid golden synthesis references and infleunce quality aggregates; remaining 26 contribute to Coverage.

Evaluation covers 32 models spanning major provider families active as of early 2026, selected  as follows. \emph{Usage-driven}: top general-purpose models by OpenRouter token consumption, including DeepSeek-V3.2, Kimi-K2.5, MiniMax-M2.5, and Gemini-2.0-Flash. \emph{Longitudinal}: every non-deprecated OpenAI model from GPT-4o through the full GPT-5 family, enabling within-provider scaling analysis. \emph{Frontier}: the full Anthropic Claude family, Google Gemini-2.5-Pro through Gemini-3-Pro, and open-weight representatives such as Qwen3.5-397B-OSS, LLaMA-4-Maverick, Mistral-Nemo, and Doubao-Seed16. All models are accessed via OpenRouter, a unified API platform providing standardized access to commercial and open-weight LLMs, under default provider hyperparameters with no per-model tuning. 
All API calls were made between February and March 2026. Each model is prompted with the original prompt verbatim, using VerilogEval's module-completion prompt or RTLLM's natural-language spec in a single-turn, zero-shot.

\subsection{Evaluation Pipeline}

Each model--task pair receives $K=5$ independent generation attempts. A generated design must clear three sequential gates before its hardware quality is assessed: \emph{syntax validity} (Icarus Verilog parse succeeds), \emph{synthesizability} (Yosys elaboration and mapping under the Nangate45 45\,nm standard-cell library completes without error), and \emph{functional correctness} (testbench simulation passes). All designs -- model-generated or golden reference-- traverse the same toolchain version and configuration. Two design decisions ensure a sound evaluation. First, simulation verdicts use fail-first matching: a pass requires an explicit pass string with no preceding failure signal, preventing false positives from partially-correct outputs. Second, RTLLM tasks  embed the target module after helper submodule definitions, so the top-module resolution step strips comments and matches task-derived module name before falling back to the first syntactic \texttt{module} keyword; naive first-keyword selection would produce spurious synthesis failures that are evaluator errors rather than model errors.
Designs that pass Icarus but fail Yosys are synthesis failures, classified into a nine-class taxonomy by  pattern matching on Yosys diagnostic output covering undefined module references, non- synthesizable constructs, elaboration-time syntax errors, and synthesis timeouts. Per-task inference metadata (API cost, reasoning token count, output throughput, and time-to-first-token) is recorded via OpenRouter. To confirm that HQI ranks generalize beyond Nangate45, all passing designs are re-synthesized under  IHP SG13G2 (130\,nm) and OSU 0.35\,$\mu$m tech libraries.

\subsection{H/W Quality Index \& Aggregate Metrics}
\paragraph{Per-attempt HQI.}
A design that fails any pipeline gate receives $\mathrm{HQI}=0$. Tasks without a golden synthesis reference (26 of 202) are assigned $\mathrm{HQI}=\mathrm{NaN}$ and excluded from quality aggregates while contributing to coverage. For a passing design on task $t$, let $\hat{a},\hat{d},\hat{w}$ be its post-synthesis area, delay, warning count, and $a^*_t,d^*_t,w^*_t$ golden reference values under the same flow. Normalized cost is

\[\mathrm{cost} = 0.5\cdot\frac{\hat{a}}{a^*_t} + 0.5\cdot\frac{\hat{d}}{d^*_t} + 0.1\cdot\max\!\left(0,\,\hat{w}-w^*_t\right),\]
yielding $\mathrm{HQI}=\min(100/\mathrm{cost},\,100)$.A score of 100 indicates parity with the reference; the cap prevents designs that may beat the golden on one metric from inflating scores. No model's Global HQI exceeds 85.1, so the ceiling does not compress rankings. Equal weights for area and delay reflect standard QoR practice; smaller warning coefficient treats warnings as a soft signal and penalizes \emph{excess} warnings above the baseline. The closed form admits application-specific reweighting 
we report equal-weight HQI as a constraint-agnostic default. Across ten weight configurations (area-only, delay-only, $3\times$ warning penalty) and equal-weight aggregation across tasks, rankings are preserved at Spearman $\rho \geq 0.985$ with rank displacement $\leq 3$ positions. 

Let $\mathcal{T}^+_m$ be tasks solved by model $m$ in at least one of five attempts, and $\mathcal{T}^\dagger$ ($|\mathcal{T}^\dagger|=176$) the tasks with valid golden references. Complexity-weighted Coverage, Global HQI (best-of-5 capability ceiling), and Expected HQI (single-attempt quality) are:
\begin{align*}
\mathrm{Coverage}(m) &= \frac{\sum_{t\in\mathcal{T}^+_m}C_t}{\sum_{t\in\mathcal{T}}C_t}\times100\%,\\[4pt]
\mathrm{GlobalHQI}(m) &= \frac{\sum_{t\in\mathcal{T}^\dagger}C_t\cdot\max_k\mathrm{HQI}_{m,t,k}}{\sum_{t\in\mathcal{T}^\dagger}C_t},\\[4pt]
\mathrm{ExpHQI}(m) &= \frac{\sum_{t\in\mathcal{T}^\dagger}C_t\cdot\bar{\mathrm{HQI}}_{m,t}}{\sum_{t\in\mathcal{T}^\dagger}C_t}.
\end{align*}
Gap between Global  and Expected HQI quantifies deployment-readiness: a large gap indicates that the model's capability ceiling cannot be realized in a single-shot that agentic pipelines demand.

\section{Results}\label{sec:results}
\subsection{Overall Landscape}

Table~\ref{tab:main_results} reports Coverage, Global HQI, and Expected HQI for all 32 models. Figure~\ref{fig:teaser} plots the same data on the Coverage--Global HQI plane. K-means clustering ($k{=}3$, silhouette $= 0.59$) identifies two gaps that define three tiers: 3.7 points between rank~14 (68.0) and rank~15 (64.3), and 7.0 points between rank~29 (46.6) and rank~30 (39.6). These rankings are robust to HQI coefficient and complexity-weighting choices (Spearman $\rho \geq 0.985$ across all perturbations).

\begin{table*}[t]
\centering
\caption{Global HQI ranks (tier-boundaries from k-means, $k{=}3$). Coverage (\%) is complexity-weighted task-solve rate; G.HQI is best-of-5 capability; E.HQI is per-attempt quality (0--100, complexity-weighted over 176 reference tasks). ~$^\dagger$ Open-weight model.}
\vspace*{-0.1in}
\label{tab:main_results}
\setlength{\tabcolsep}{2pt}\small
\resizebox{\linewidth}{!}{%
\begin{tabular}{@{}lrrr@{\hspace{12pt}}lrrr@{\hspace{12pt}}lrrr@{}}
\toprule
\multicolumn{4}{c}{Tier 1 (G.HQI $\geq$66, 14 models)} &
\multicolumn{4}{c}{Tier 2 (G.HQI 43--66, 15 models)} &
\multicolumn{4}{c}{Tier 3 (G.HQI $<$43, 3 models)} \\
\cmidrule(r){1-4}\cmidrule(lr){5-8}\cmidrule(l){9-12}
Model & Cov. & G.HQI & E.HQI & Model & Cov. & G.HQI & E.HQI & Model & Cov. & G.HQI & E.HQI \\
\midrule
Gemini-3-Pro        & 87.5 & 85.1 & 78.5 & Gemini-2.5-Pro       & 63.7 & 64.3 & 56.0 & Gemini-2.0-Flash    & 40.9 & 39.6 & 28.6 \\
GPT-5.4-Pro         & 84.5 & 81.3 & 77.5 & MiniMax-M2.5$^\dagger$  & 60.3 & 63.0 & 49.7 & GPT-5-Nano          & 36.8 & 35.0 & 19.9 \\
Gemini-3-Flash      & 83.8 & 81.2 & 70.0 & GPT-4.1              & 66.7 & 62.8 & 48.5 & Mistral-Nemo$^\dagger$ & 22.3 & 18.1 &  9.4 \\
GPT-5.3-Codex       & 83.8 & 80.8 & 75.7 & GLM-5$^\dagger$         & 62.5 & 62.6 & 50.4 &                     &      &      &      \\
GPT-5-Pro           & 82.6 & 80.5 & 72.3 & Kimi-K2.5$^\dagger$     & 62.1 & 62.2 & 52.7 &                     &      &      &      \\
Claude-4.6-Sonnet   & 82.9 & 78.2 & 71.0 & Qwen3.5-Plus         & 60.0 & 60.3 & 47.0 &                     &      &      &      \\
Claude-4.6-Opus     & 81.4 & 78.0 & 74.3 & DeepSeek-V3.2$^\dagger$ & 59.4 & 58.8 & 46.2 &                     &      &      &      \\
GPT-5.4             & 75.8 & 74.7 & 67.5 & GPT-OSS-120B$^\dagger$  & 56.3 & 58.8 & 40.9 &                     &      &      &      \\
GPT-5.1             & 77.0 & 73.5 & 60.5 & Qwen3.5-397B$^\dagger$  & 55.2 & 58.2 & 47.5 &                     &      &      &      \\
Claude-3.5-Sonnet   & 75.1 & 73.1 & 62.9 & Doubao-Seed16$^\dagger$ & 51.1 & 53.3 & 42.7 &                     &      &      &      \\
Claude-3.7-Sonnet   & 75.0 & 72.3 & 62.4 & GPT-5               & 48.2 & 50.5 & 35.3 &                     &      &      &      \\
GPT-5.2             & 72.7 & 72.2 & 63.1 & LLaMA-4-Mav.$^\dagger$ & 50.0 & 48.8 & 38.0 &                     &      &      &      \\
Claude-4.5-Sonnet   & 73.8 & 71.6 & 62.4 & Grok-4.1-Fast       & 47.3 & 47.5 & 36.0 &                     &      &      &      \\
GPT-4o              & 71.2 & 68.0 & 50.2 & Gemini-2.5-Flash    & 46.6 & 47.3 & 33.2 &                     &      &      &      \\
                    &      &      &      & GPT-5-Mini          & 49.3 & 46.6 & 24.5 &                     &      &      &      \\
\bottomrule
\end{tabular}}
\end{table*}

\textbf{Tier~1} (Global HQI $\geq 66$, 14 models) forms a distinct frontier cluster led by Gemini-3-Pro (87.5\% coverage, 85.1 Global HQI), with the four next-ranked models all exceeding 80; the tier spans 71.2--87.5\% coverage and includes all five Claude variants, seven OpenAI models from GPT-4o through GPT-5.4-Pro, and two Google models. \textbf{Tier~2} (Global HQI 43--66, 15 models) forms a dense cluster visible in the inset, including Gemini-2.5-Pro (64.3) and the strongest open-weight models, with DeepSeek-V3.2 at 58.8 and Qwen3.5-397B-OSS at 58.2; the open-weight frontier trails the lowest Tier~1 model by roughly 10 Global HQI points. GPT-5 base (50.5) places in Tier~2 while later variants in the same family reach Tier~1, indicating that post-training refinement, rather than pre-training scale alone, determines RTL capability. \textbf{Tier~3} (Global HQI $< 43$, 3 models) contains Gemini-2.0-Flash (39.6), GPT-5-Nano (35.0), and Mistral-Nemo (18.1), all scoring well below the Tier~2 floor.

\begin{figure}[t]
  \centering
  \includegraphics[width=\linewidth]{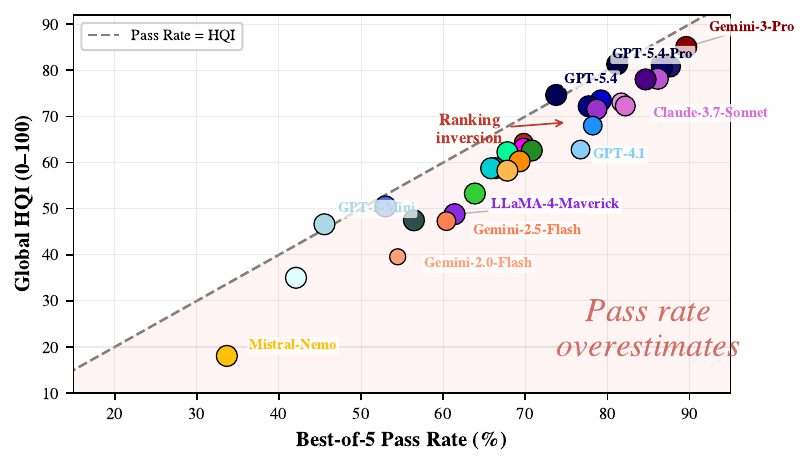}
  \Description{Scatter plot of 32 models on the Best-of-5 Pass Rate versus Global HQI plane. A dashed diagonal line marks parity (pass rate equals HQI). Nearly all models fall below the diagonal, indicating that pass rate systematically overestimates hardware quality. Key outliers are labeled: GPT-4.1 and Gemini-2.0-Flash show large gaps (13.9 and 14.9 points respectively), while GPT-5.4-Pro nearly touches the diagonal. An annotation highlights the ranking inversion between GPT-4.1 and GPT-5.4.}
   \vspace*{-0.15in}
  \caption{Best-of-5  vs.\ HQI. Points below diagonal show pass rate overestimates implementation quality; 29 models are there. Mean gap is 7.5 pts, rank inversions of upto 7 positions.}
  \label{fig:passrate_vs_hqi}
\end{figure}

Simulation-based pass rate systematically overstates hardware readiness (Figure~\ref{fig:passrate_vs_hqi}): best-of-five pass rate exceeds Global HQI by 7.5 points on average, with gaps reaching 13.9 (GPT-4.1) and 14.9 (Gemini-2.0-Flash) for models that solve many tasks at mediocre quality. Pass rate also distorts rankings: GPT-4.1 ranks 14th by pass rate but 17th by HQI, while GPT-5.4-Pro shifts 9 $\rightarrow$2 because its passing designs closely match expert quality, so a pass-rate-only benchmark inverts GPT-4.1 vs.\ GPT-5.4 ordering despite GPT-5.4 producing better hardware.

\subsection{Within-Provider Scaling: OpenAI Lineage}
OpenAI lineage from GPT-4o$\rightarrow$ GPT-5  reveals two patterns. Point releases trend upward but non-monotonically: GPT-5.1 (73.5) $\rightarrow$ GPT-5.2 (72.2) $\rightarrow$ GPT-5.3-Codex (80.8) $\rightarrow$ GPT-5.4 (74.7) $\rightarrow$ GPT-5.4-Pro (81.3), with more gains in Pro and Codex variants. Generational transitions carry risk: GPT-5 base underperforms GPT-4o (50.5 vs.\ 68.0) by 17.5 points that is recovered after many point releases. Model capacity within a generation matters: GPT-5-Pro (80.5) outperforms GPT-5-Nano (35.0) by $>$ 45 HQI points, more than software benchmark separations; RTL ability is shaped by post-training optimization and model capacity. 

\subsection{Reliability: Best-of-5 vs Per Attempt}
The gap between Global and Expected HQI does not diminish. 
In Tier~1, the gap ranges from 3.7 to 17.8 points (median 8.7); even the top model drops 6.6 HQI points from its best-of-five ceiling to its single-attempt. In Tier~2 and Tier~3, several models show gaps exceeding 15 points, with the largest reaching 22.1. A single call may retain half the quality that five independent attempts would achieve. For agentic pipelines that propagate the first attempt unsupervised, these gaps reveal a deployment-readiness deficit; closing it requires improving single-shot reliability.

\subsection{Category-Level Analysis}

\begin{figure*}[t]
  \centering
  \includegraphics[width=\linewidth]{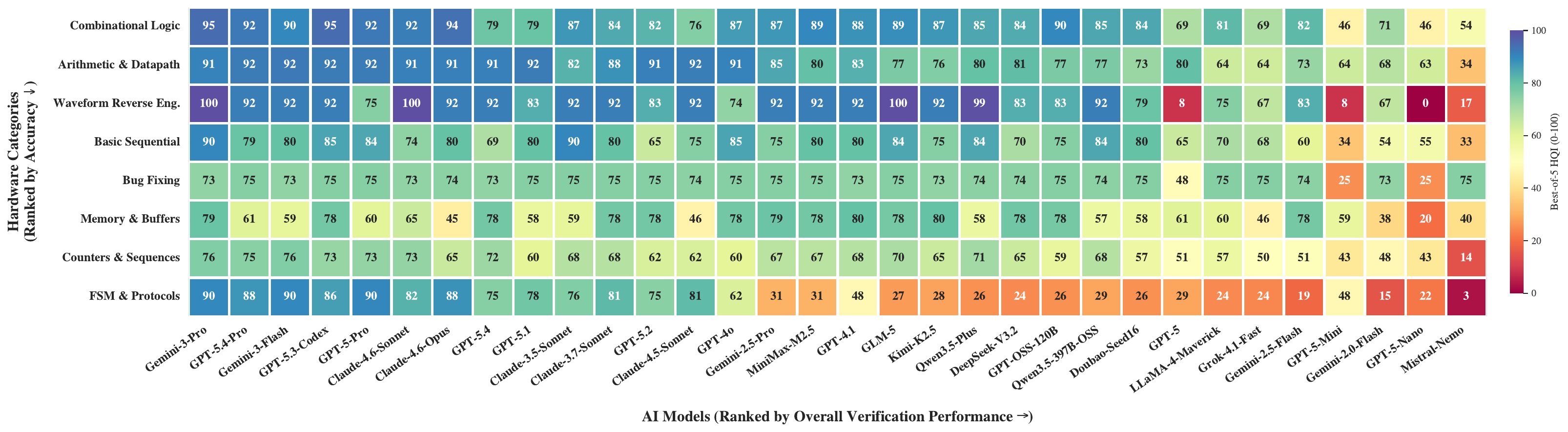}
  \Description{A color-coded heatmap with 32 AI models on the horizontal axis ordered left-to-right by Global HQI (Gemini-3-Pro to Mistral-Nemo) and eight hardware design categories on the vertical axis ordered top-to-bottom by average Tier-1 accuracy (Combinational Logic easiest, FSM and Protocols hardest). Each cell shows the best-of-five HQI score (0--100); blue cells indicate high scores and orange-red cells indicate low scores. Tier-1 frontier models achieve scores of 87--95 on Combinational Logic but drop to 19--90 on FSM and Protocols. Tier-3 models frequently show orange-red cells on FSM and Memory categories.}
  \vspace*{-0.15in}
  \caption{Best-of-five HQI heatmap across eight hardware categories and 32 models. Models are ordered left-to-right by Global HQI; categories are ordered top-to-bottom by average Tier~1 accuracy. Color encodes HQI score (0--100).}
  \label{fig:heatmap_best5}
\end{figure*}

\begin{figure*}[t]
  \centering
  \includegraphics[width=\linewidth]{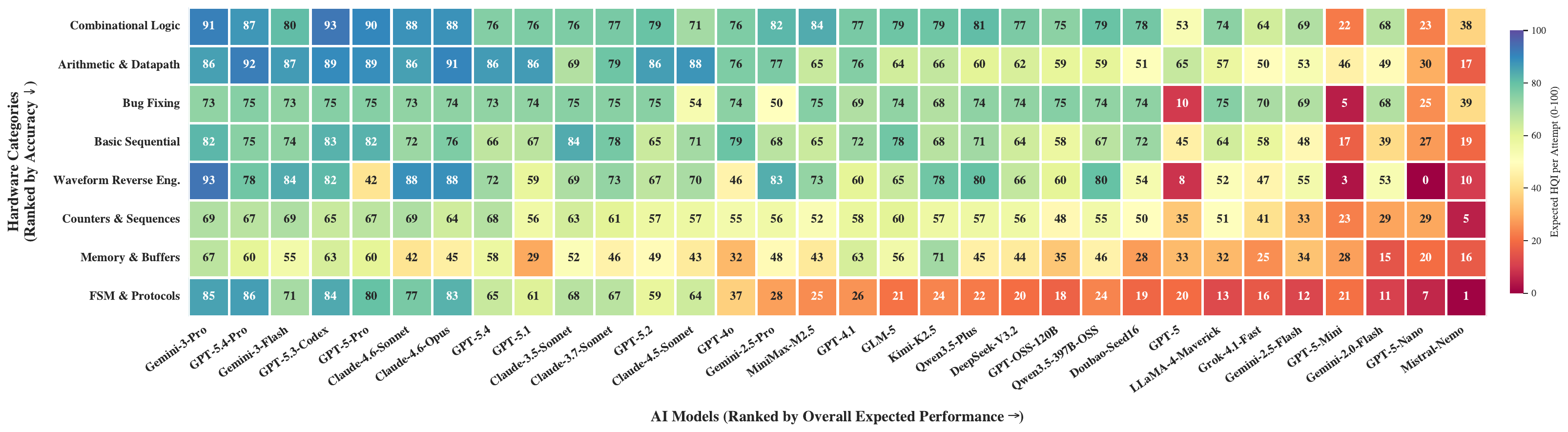}
  \Description{A color-coded heatmap identical in layout to the best-of-five heatmap but showing per-attempt Expected HQI scores. The color pattern shifts toward lower values across all models and categories relative to the best-of-five heatmap, reflecting single-attempt degradation. The most pronounced drops appear in FSM and Protocols and Memory and Buffers. Tier-1 models that scored above 80 on best-of-five commonly drop to 60--75 on per-attempt in the hardest categories.}
\vspace*{-0.25in}
  \caption{Per-attempt Expected HQI heatmap across 8 h/w categories and 32 models, for same ordering as Figure~\ref{fig:heatmap_best5}. Scores reflect single-attempt quality. 
  }
  \label{fig:heatmap_perattempt}
\end{figure*}

Figure~\ref{fig:heatmap_best5} and Figure~\ref{fig:heatmap_perattempt} present best-of-five and per-attempt HQI heatmaps across all 32 models and eight task categories. \textbf{Combinational Logic} is the easiest category: Tier~1 models achieve 87--95 and even the weakest models exceed 40. \textbf{FSM \& Protocols} and \textbf{Memory \& Buffers} are hardest, with high variance and Tier~3 models scoring near zero. \textbf{Waveform Reverse Engineering} is bimodal: Tier~1 and Tier~2 models  exceed 75, while Tier~3 models score zero. The per-attempt heatmap uniformly shifts downward relative to best-of-five, with largest drops in the hardest categories.

\begin{figure*}[t]
\centering
\begin{minipage}[b]{0.33\linewidth}
  \centering
  \includegraphics[height=2.8in]{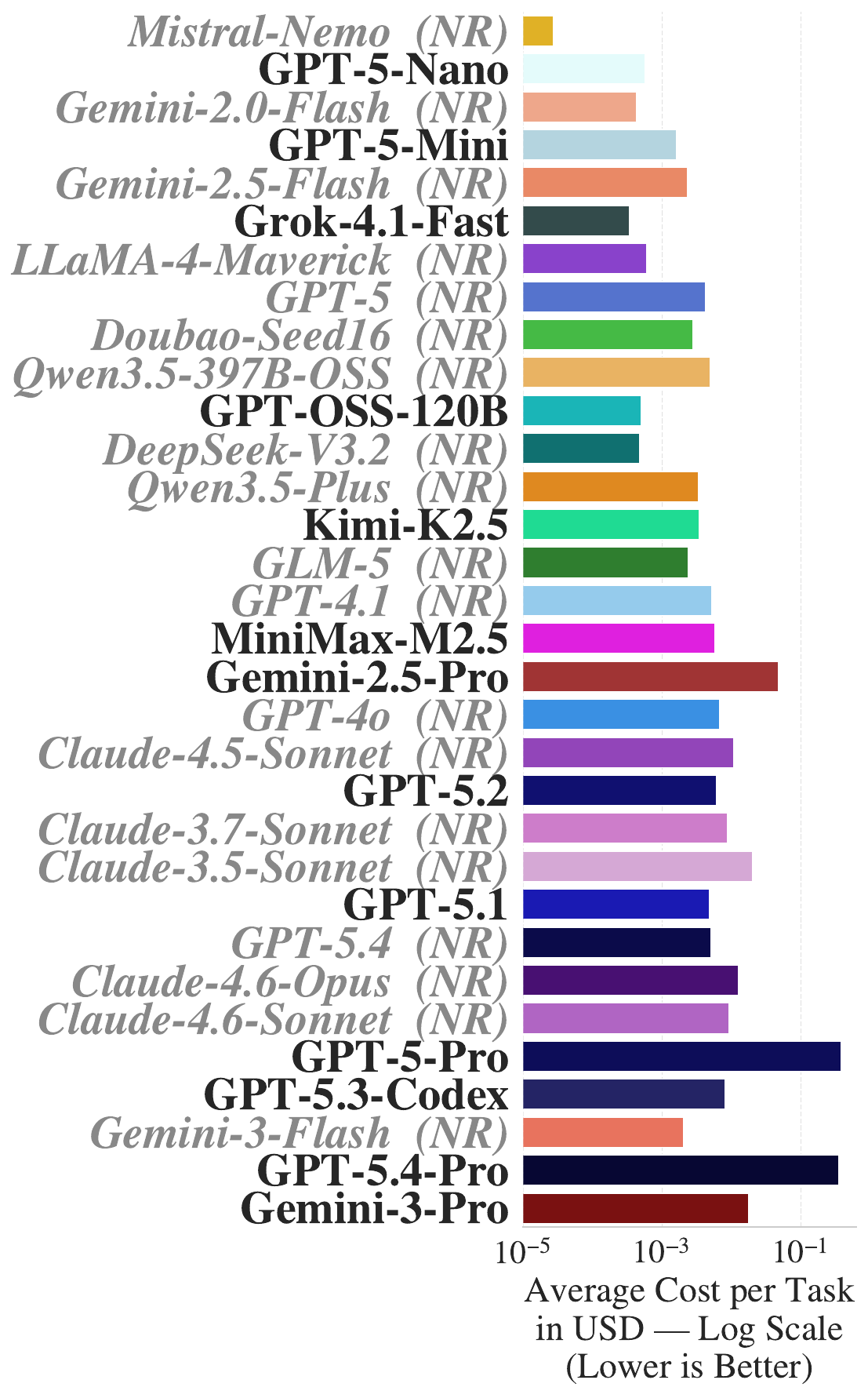}\\[2pt]
  \small (a) Cost (USD, log-scale, lower is better).
\end{minipage}%
\hfill
\begin{minipage}[b]{0.155\linewidth}
  \centering
  \includegraphics[height=2.8in]{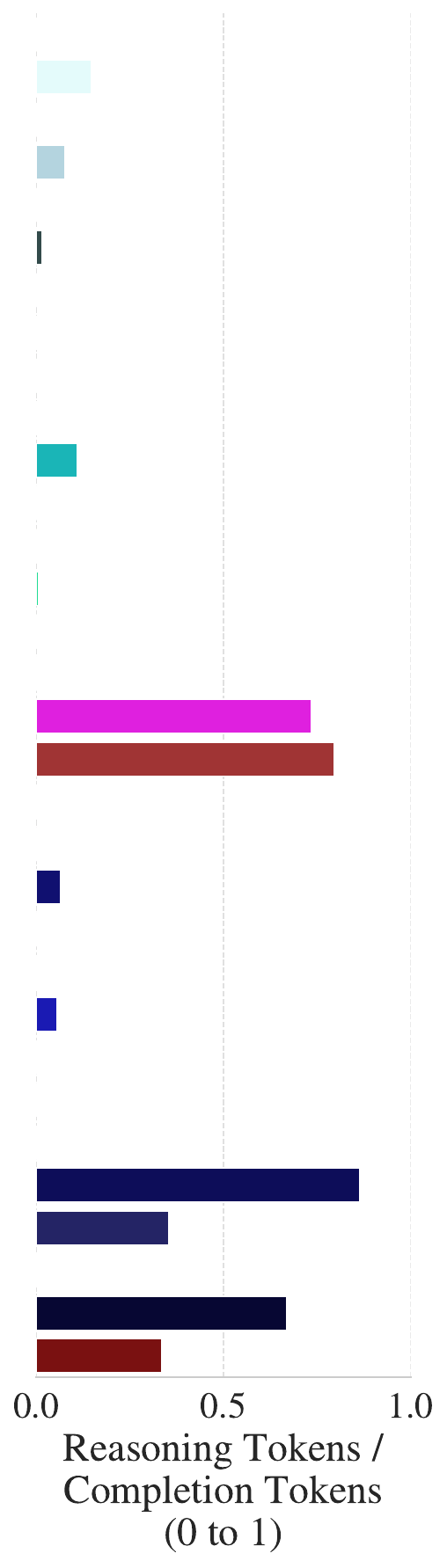}\\[2pt]
  \small (b) Reasoning ratio.
\end{minipage}%
\hfill
\begin{minipage}[b]{0.155\linewidth}
  \centering
  \includegraphics[height=2.8in]{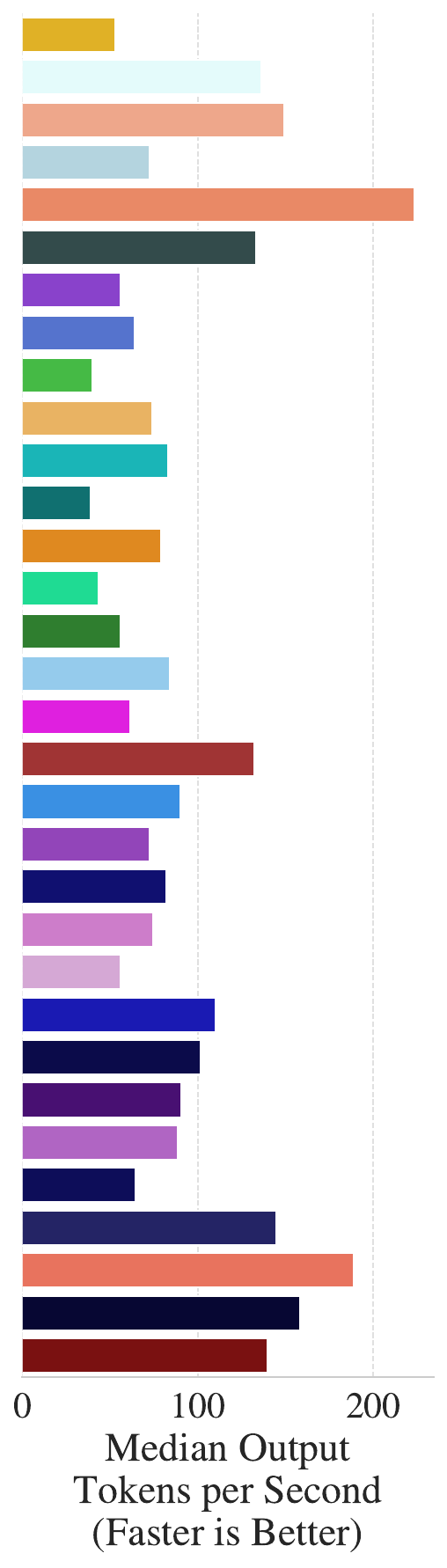}\\[2pt]
  \small (c) Throughput (tok/s).
\end{minipage}%
\hfill
\begin{minipage}[b]{0.155\linewidth}
  \centering
  \includegraphics[height=2.8in]{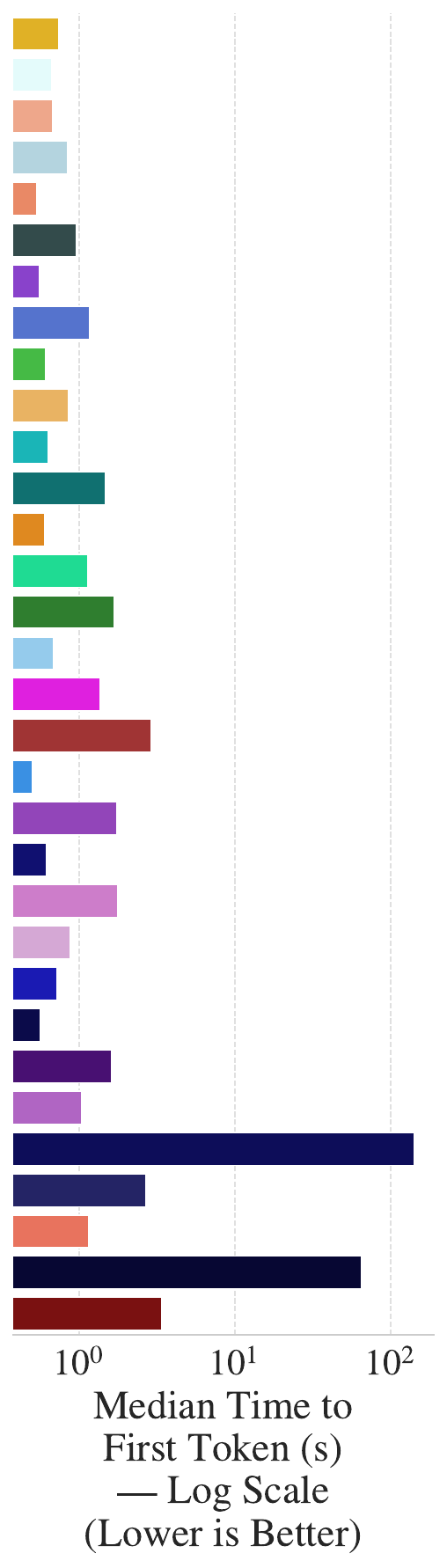}\\[2pt]
  \small (d) TTFT (s, log-scale).
\end{minipage}%
\hfill
\begin{minipage}[b]{0.155\linewidth}
  \centering
  \includegraphics[height=2.8in]{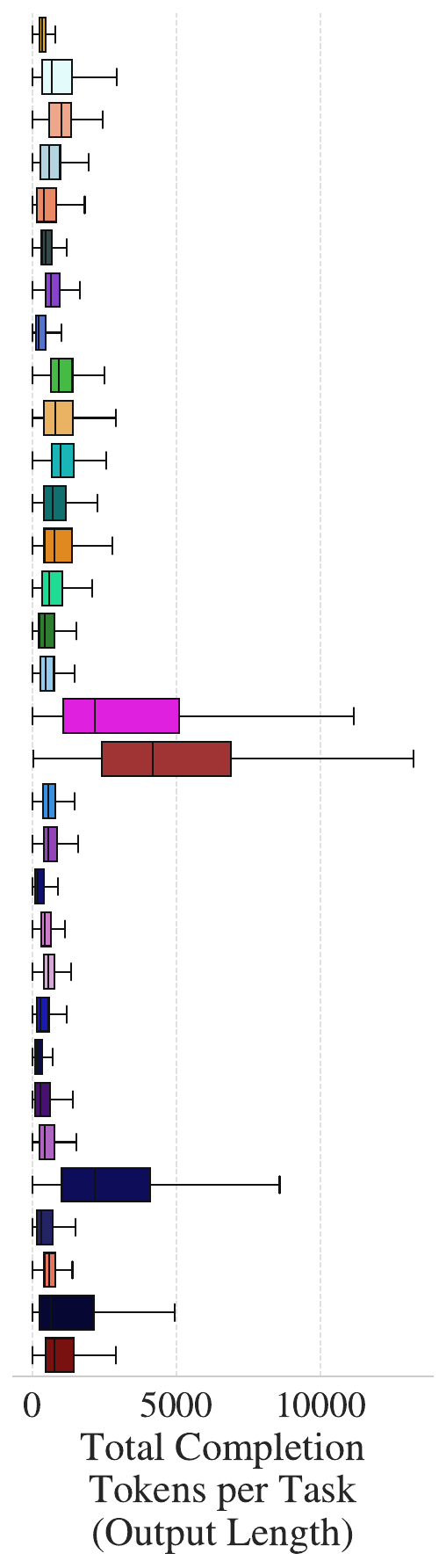}\\[2pt]
  \small (e) Verbosity (tokens).
\end{minipage}
\Description{A row of five narrow vertical bar charts profiling all 32 models on five inference dimensions. (a) Per-task API cost on log scale; GPT-5-Pro and GPT-5.4-Pro are most expensive, Mistral-Nemo cheapest. (b) Reasoning token ratio from 0 to 1; GPT-5-Pro highest at 0.88, Gemini-2.5-Pro at 0.80; 19 models show zero ratio and are labeled Non-Reasoning LLM. (c) Output throughput in tokens per second; Gemini-2.5-Flash leads at 215. (d) TTFT latency on log scale in seconds; GPT-4o fastest near 0.5 seconds, GPT-5-Pro slowest at over 100 seconds. (e) Box plot of total completion tokens per task; Gemini-2.5-Pro most verbose with median near 4000, GPT-5.4 least verbose.}
\caption{Inference characteristics across all models measured via the OpenRouter API. Cost and TTFT use log scale.}
\vspace*{-0.15in}
\label{fig:inference}
\end{figure*}
\subsection{Synthesis Failure Analysis}

Among the 32{,}320  attempts, 195 exhibit  synthesis failures (designs that pass the Icarus Verilog parser but fail Yosys elaboration or synthesis). Tool-based pattern matching on Yosys diagnostic output yields a detailed taxonomy; three dominant modes account for 76.6\% of all failures. \emph{Late syntax errors} (59 tags, 30.0\%) arise when code passes Verilog parser but is rejected at Yosys elaboration due to malformed always-blocks, misplaced begin/end delimiters, or structural issues missed by the front-end parser; 55-of-59 cases originate from RTLLM tasks, reflecting the structural complexity of multi-module designs. \emph{Undefined module references} (50 tags, 25.4\%) occur on VerilogEval tasks where models omit the \texttt{top\_module} wrapper. \emph{Non-synthesizable constructs} (41 tags, 20.8\%) include while loops with non-constant bounds, non-synthesizable initial blocks, and other behavioral constructs valid in simulation but unsupported by synthesis, mostly on RTLLM tasks (31 of 41). Rest are less frequent yet  informative: \emph{simulation-only system tasks} (14 tags, 7.1\%), such as \texttt{\$finish} and \texttt{\$display} embedded in synthesized module bodies, indicate confusion between testbench and design code; \emph{synthesis timeout} (14 tags, 7.1\%) on VerilogEval tasks and for proprietary models; pathologically large or nested generated netlists exhaust 30-second synthesis budget; and \emph{invalid nesting} (11 tags, 5.6\%) covers illegal structural constructs such as nested always-blocks.  Benchmark-level split confirms that taxonomy captures task-format differences: RTLLM failures are in late syntax errors and non-synthesizable constructs (84\%), VerilogEval failures cluster on undefined module references and synthesis timeout (69\%). 

The taxonomy reveals a systematic split. \textbf{Proprietary models} (103 failure tags) fail \emph{late}: their dominant mode is late syntax error (46\%), where code passes the parser but fails at Yosys elaboration, and synthesis timeout (12\%) is exclusive to this group, arising from complex generated netlists. This pattern is consistent with exposure to synthesis-grade Verilog, where structurally plausible RTL reaches the elaborator before failing. \textbf{Open-weight models} (92 failure tags) fail \emph{early} through RTL discipline violations: undefined module reference (40\%), non-synthesizable constructs (29\%), simulation-only system tasks (13\%) jointly account for 82\% of open-weight failures, consistent with training on simulation-oriented Verilog rather than synthesis-grade RTL.  $19\times$ ratio in failure count between the weakest and strongest models, combined with this shift in failure mode, is consistent with differences in training data composition; curation of synthesis-grade RTL corpora can be a lever to close this gap, though our observational data does not imply causation.

Individual models exhibit unique failure profiles. Mistral-Nemo accounts for 57 of 195  failures (29\%), mostly \emph{undefined module} and \emph{non-synthesizable} tags. Among frontier models, Claude-4.5-Sonnet fails as synthesis timeouts (4 of 5, complex netlists) and Claude-3.7-Sonnet are \emph{invalid nesting}; GPT-5.4 and -Pro each incur three failures despite solving over 80\% of tasks. Fine-tuning on failure modes (module wrapper conventions for open-weight models, netlist complexity control for Claude) may improve. Re-synthesizing passing designs with IHP SG13G2 (130\,nm) and OSU 0.35\,$\mu$m yields Spearman $\rho > 0.99$ agreement with Nangate45 ranks.

\subsection{Inference Characteristics}
Figure~\ref{fig:inference} profiles models across five deployment dimensions via OpenRouter API. Cost and quality poorly correlate, heavy reasoning overhead does not guarantee RTL quality, and deployment choices depend on operative constraint. Per-task API cost spans five orders of magnitude (Figure~\ref{fig:inference}(a)), yet expensive models are not the best. GPT-5-Pro (\$0.26/task, 80.5 HQI) and GPT-5.4-Pro (\$0.12/task, 81.3) are \emph{Pareto-dominated}: Gemini-3-Pro achieves 85.1 HQI at \$0.01/task. Cost-efficiency frontier haes Gemini-3-Flash (\$0.002, 81.2 HQI), GPT-5.3-Codex (\$0.005, 80.8), Gemini-3-Pro (\$0.01, 85.1), delivering Tier~1 quality at 10--130$\times$ lower cost.
Reasoning does not guarantee superior RTL quality. GPT-5-Pro devotes 90\% of output to reasoning (Figure~\ref{fig:inference}(b)), pushing TTFT to 142\,s and throughput to 12\,tok/s, yet scores below non-reasoning Gemini-3-Flash (81.2 HQI, 1.2\,s TTFT, \$0.002/task). 7-of-14 Tier~1 models are non-reasoning; for cost-sensitive batch generation, five attempts of Gemini-3-Flash cost less than a GPT-5-Pro call while delivering higher best-of-5 quality. Output verbosity (Figure~\ref{fig:inference}(e)) is uncorrelated with quality, confirming RTL quality depends on precision, not volume.

\section{Conclusion}
This paper introduced a synthesis-in-the-loop evaluation framework covering 32 models, 202 tasks, and five independent attempts each, and showed that functional correctness alone captures neither the capability ceiling nor the deployment reliability of current models. HQI reveals a  three-tier landscape in which simulation-based pass rates overstate hardware readiness: models can pass 77\% of tasks while achieving only 63 Global HQI, confirming that synthesis-aware metrics are necessary for RTL evaluation. Three findings have practical implications. The best-of-five-to-single-attempt gap (3.7--22.1 HQI points) shows that they cannot yet be integrated in an agentic framework; narrowing this gap by improving single-shot reliability is the path toward autonomous pipelines. Within-generation model capacity matters disproportionately for RTL (45+ HQI-point spread within the GPT-5 family), and generational transitions can regress before point releases recover. The third finding is a systematic divergence in failure mode: proprietary models fail late (elaboration errors, synthesis timeout) while open-weight models fail early through structural RTL  violations, a pattern consistent with differences in training data composition; curation of synthesis-grade training data is an improvement lever. 

{\it Limitations and future work.}
This evaluation uses single turn, zero-shot prompting. Iterative refinement with synthesis feedback ~\cite{liu2025rtlcoder} may shrink capability gaps. Since VerilogEval and RTLLM are open benchmarks, data contamination is not ruled out~\cite{wang2025vericontaminated}: a memorized solution can synthesize efficiently, so HQI does not certify novelty; a held-out task set offers the strongest guarantee. Synthesis flow uses Yosys with open standard-cell library; area and delay numbers will differ for industrial flows (Cadence, Synopsys), although Spearman $\rho > 0.99$ across three technology libraries suggests relative orderings transfer. The pipeline assesses synthesizability and post-synthesis QoR but does not exercise lint, clock-domain-crossing, reset-quality, design-for-test, or multi-mode multi-corner timing closure. Place-and-route metrics such as congestion and hold-time violations may further differentiate models. The task set excludes system-level integration, analog/mixed-signal, floorplan-aware design, and benchmark is a snapshot whose rankings will evolve with model updates.

\begin{acks}
This work was supported in part by the Google Ph.D. Fellowship, the NYUAD Center for Cyber Security (CCS), funded by Tamkeen under the NYUAD Research Institute Award G1104, and by the National Science Foundation under awards 2340949 and 2419880. This research was carried out on the High Performance Computing resources at NYUAD and at Kansas State University.
\end{acks}

\bibliographystyle{ACM-Reference-Format}
\bibliography{ref}

\end{document}